\documentclass[prl,aps,amsfonts,amssymb,amsmath,floats,floatfix,showpacs,twocolumn,superscriptaddress,unsortedaddress,letterpaper]{revtex4}
\usepackage{graphics}
\usepackage{epsfig,graphicx}

\begin{document}

\title{Mott versus Slater-type metal-insulator transition in Mn-substituted Sr$_3$Ru$_2$O$_7$}

\author{M.A.\,Hossain}
\email{hossain@slac.stanford.edu} 
\affiliation{Department of Physics {\rm {\&}} Astronomy, University of British Columbia, Vancouver, British Columbia V6T\,1Z1, Canada} 
\affiliation{Advanced Light Source, Lawrence Berkeley National Laboratory, Berkeley, California 94720, USA}
\author{B.\,Bohnenbuck}
\affiliation{Max-Planck-Institut f\"{u}r Festk\"{o}rperforschung, Heisenbergstra\ss e 1, 70569 Stuttgart, Germany}
\author{Y.D.\,Chuang}
\affiliation{Advanced Light Source, Lawrence Berkeley National Laboratory, Berkeley, California 94720, USA}
\author{M.W.\,Haverkort}
\affiliation{Max-Planck-Institut f\"{u}r Festk\"{o}rperforschung, Heisenbergstra\ss e 1, 70569 Stuttgart, Germany}
\author{I.S.\,Elfimov}
\affiliation{Department of Physics {\rm {\&}} Astronomy, University of British Columbia, Vancouver, British Columbia V6T\,1Z1, Canada}
\affiliation{Quantum Matter Institute, University of British Columbia, Vancouver, British Columbia V6T\,1Z4, Canada}
\author{A.\,Tanaka}
\affiliation{Department of Quantum Matter, ADSM, Hiroshima University, Higashi-Hiroshima 739-8530, Japan}
\author{A.G.\,Cruz}
\affiliation{Advanced Light Source, Lawrence Berkeley National Laboratory, Berkeley, California 94720, USA}
\author{Z.\,Hu}
\affiliation{Max Planck Institute for Chemical Physics of Solids, 01187 Dresden, Germany}
\author{H.-J. Lin}
\affiliation{National Synchrotron Radiation Research Center, 101 Hsin-Ann Road, Hsinchu 30077, Taiwan}
\author{C.T. Chen}
\affiliation{National Synchrotron Radiation Research Center, 101 Hsin-Ann Road, Hsinchu 30077, Taiwan}
\author{R.\,Mathieu}
\affiliation{Department of Applied Physics, University of Tokyo, Tokyo 113-8656, Japan}
\author{Y.\,Tokura}
\affiliation{Department of Applied Physics, University of Tokyo, Tokyo 113-8656, Japan}
\author{Y.\,Yoshida}
\affiliation{National Institute of Advanced Industrial Science and Technology  (AIST), Tsukuba, 305-8568, Japan}
\author{L.H.\,Tjeng}
\affiliation{Max Planck Institute for Chemical Physics of Solids, 01187 Dresden, Germany}
\author{Z.\,Hussain}
\affiliation{Advanced Light Source, Lawrence Berkeley National Laboratory, Berkeley, California 94720, USA}
\author{B.\,Keimer}
\affiliation{Max-Planck-Institut f\"{u}r Festk\"{o}rperforschung, Heisenbergstra\ss e 1, 70569 Stuttgart, Germany}
\author{G.A.\,Sawatzky}
\affiliation{Department of Physics {\rm {\&}} Astronomy, University of British Columbia, Vancouver, British Columbia V6T\,1Z1, Canada}
\affiliation{Quantum Matter Institute, University of British Columbia, Vancouver, British Columbia V6T\,1Z4, Canada}
\author{A.\,Damascelli}
\email{damascelli@physics.ubc.ca} 
\affiliation{Department of Physics {\rm {\&}} Astronomy, University of British Columbia, Vancouver, British Columbia V6T\,1Z1, Canada}
\affiliation{Quantum Matter Institute, University of British Columbia, Vancouver, British Columbia V6T\,1Z4, Canada}

\begin{abstract}
We present a temperature-dependent x-ray absorption (XAS) and resonant elastic x-ray scattering (REXS) study of the metal-insulator transition (MIT) in Sr$_3$(Ru$_{1-x}$Mn$_x$)$_2$O$_7$. The XAS results reveal that the MIT drives the onset of local antiferromagnetic correlations around the Mn impurities, a precursor of the long-range antiferromagnetism detected by REXS at $T_{order}\!<\!T_{MIT}$.  This establishes that the MIT is of the Mott-type (electronic correlations) as opposed to Slater-type (magnetic order). While this behavior is induced by Mn impurities, the $(\frac{1}{4},\frac{1}{4},0)$ order exists for a wide range of Mn concentrations, and points to an inherent instability of the parent compound.
\end{abstract}

\date{March 17, 2012}



\maketitle

Electronic and lattice instabilities in strongly correlated electron systems give rise to many fascinating phenomena, such as various types of spin, charge, and orbital ordering. This is a common feature of many of the 3$d$ transition-metal oxides, with the best-known examples including the stripe instability in the cuprate superconductors and the magnetic phase-separation in manganites. Competing instabilities and ordering phenomena can also be found in the somewhat less correlated 4$d$ transition-metal oxides, with the ruthenates being one of the most prominent families. Sr$_3$Ru$_2$O$_7$ is known as a metamagnetic metal on the verge of ferromagnetism \cite{ikeda}, exhibiting a magnetic field tuned quantum criticality \cite{grigera} and electronic nematic fluid behavior \cite{borzi}. However, the deeper connection between these effects is still highly debated and its description will depend on a fuller understanding of the incipient instabilities in Sr$_3$Ru$_2$O$_7$.

Magnetic impurities such as Mn have been introduced in Sr$_3$Ru$_2$O$_7$ in the attempt to stabilize the magnetic tendencies \cite{mathieu}. It was shown that due to the interplay between localized Mn\,3$d$ and delocalized Ru\,4$d$\,-\,O\,2$p$ valence states in Sr$_3$(Ru$_{1-x}$Mn$_x$)$_2$O$_7$, the Mn impurities display a ${3+}$ oxidation state and a crystal field level inversion already at room temperature \cite{hossain_PRL,Panaccione_NJP}. Upon lowering the temperature, a metal-insulator transition (MIT) has been observed for 5\,\% Mn at $T_{MIT}\!\simeq\!50$\,K, and at progressively higher $T_{MIT}$ upon increasing the Mn content \cite{mathieu}. Long-range antiferromagnetic (AF) order with a (1/4,1/4,0) magnetic wavevector was observed by neutron scattering below $T_{MIT}$ for 5\% Mn, and it was speculated to exist for all Mn concentrations up to 20\% \cite{mathieu}. Recently, however, it was proposed that a true phase transition -- seemingly associated with the onset of AF order -- is detected only at the temperature $T_M\!<\!T_{MIT}$ at which the magnetic susceptibility exhibits a maximum \cite{Hu}. This apparent disconnection between MIT and AF order reopens the question of what the origin of the MIT might be: ({\it i\,}) whether of the Mott type --  {\it electronic-correlation driven} -- with corresponding mass enhancement and formation of local moments, and only subsequently of magnetic order; ({\it ii\,}) or instead of the Slater type --  {\it magnetic order driven} -- in which a band-like insulating state with an even number of electrons is achieved via the magnetic-order-induced folding of the Brillouin zone.
\begin{figure}
\centerline{\epsfig{figure=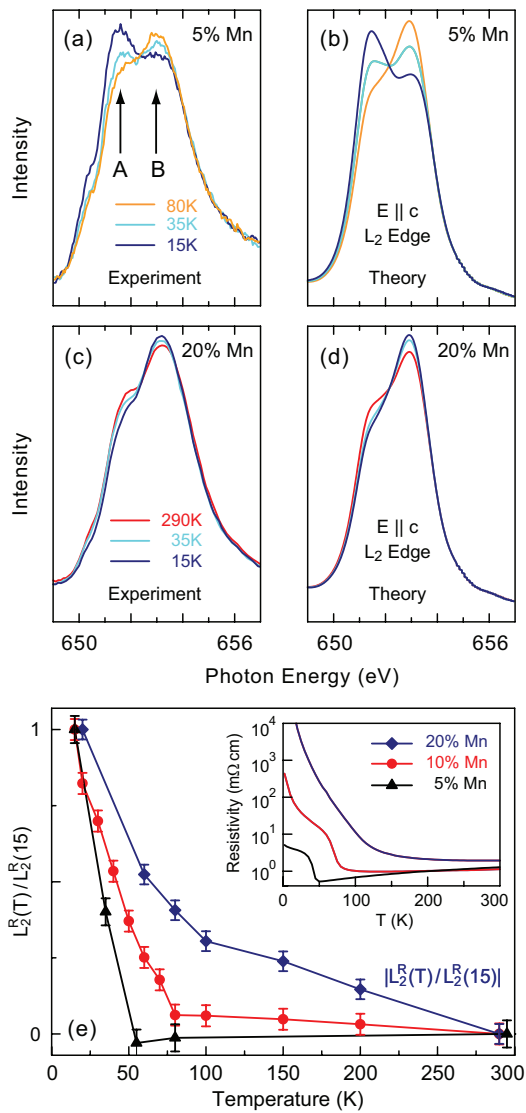,width=0.85\linewidth,clip=}} \vspace{-0.25cm}
\caption{(color online). (a,c) Temperature dependent Mn $L_2$-XAS spectra
measured on 5 and 20\% Mn-substituted Sr$_3$Ru$_2$O$_7$ with $\bf{E}\!\parallel\!c$. (b,d) Theoretical $L_2$-spectra calculated with the inclusion of a 1\,meV AF exchange field: (b) $H^{ex}$ 30$^{\circ}$ away from $c$-axis, $T_{Neel}\!=\!50$K; (d) $H^{ex}$ 60$^{\circ}$ away from $c$-axis, $T_{Neel}\!=\!100$K. (e) Normalized peak-height ratio $L_2^R(T) / L_2^R(15)$ vs. temperature for 5, 10, and 20\% Mn-substitution; $L_2^R(T)\!=\![\frac{I_A-I_B}{I_A+I_B}(T)\!-\!\frac{I_A-I_B}{I_A+I_B}(295)]$ and $I_{A,B}$ is the intensity of peak $A,B$. For 20\% Mn, the modulus of the data is plotted. Inset of (e): dc electrical resistivity data (in logarithmic scale), for the same samples \cite{mathieu}.}\label{Fig1}
\end{figure}

To understand the nature of the MIT and its interplay with magnetic correlations -- as well as the role of Mn in the context of the physics of Sr$_3$Ru$_2$O$_7$ -- we use x-ray absorption spectroscopy (XAS) and resonant elastic x-ray scattering (REXS). These techniques, owing to the element specificity of the resonant absorption processes, allow probing selectively Mn and Ru response. In addition, while XAS is sensitive to the nearest-neighbor spin-spin correlation function, and thus to {\it local AF correlations} \cite{kuiper,alders}, REXS is sensitive to spin and charge {\it long-range order} \cite{REXS}. By combining XAS and REXS we show that the MIT drives the onset of an exchange field at the Mn sites, i.e. localized AF correlations, and is only when the correlation length and/or the Mn concentration increase that a global long-range order is established. The MIT is thus of the Mott type. While both MIT and AF are induced by the Mn impurities, the (1/4,1/4,0) order exist for a wide range of Mn concentrations and points to an incipient instability of the parent compound.

XAS measurements were performed in the total electron-yield mode on the dragon beamline at NSRRC in Taiwan. REXS experiments were carried out at beamline 8.0.1 at ALS in Berkeley (Mn $L$-edges), and KMC-1 at BESSY in Berlin (Ru $L$-edges). The scattering measurements used a two-circle ultra-high-vacuum diffractometer in horizontal scattering geometry, with the incident photon beam polarized parallel to the diffraction plane ($\pi$); the scattered signal was not polarization analysed and so could contain polarization components parallel ($\pi'$) and perpendicular ($\sigma'$) to the diffraction plane. Sr$_3$(Ru$_{1-x}$Mn$_x$)$_2$O$_7$ single crystals grown by the floating zone technique \cite{mathieu} were prepared for XAS by {\it in situ} cleaving at pressures better than 2\,$\times$\,10$^{-9}$\,mbar, and cut and polished along the (110) direction for REXS. 

Fig.\,\ref{Fig1}(a,c) present temperature-dependent XAS data for 5 and 20\% Mn. Although similar behavior is observed for both $L_{2,3}$ Mn-edges, here we focus on the $L_2$-edge since it provides the highest sensitivity to the onset of local AF correlations \cite{kuiper,alders}. A pronounced evolution can be observed for the 5\% sample, with peaks A and B, respectively, gaining and loosing intensity across the MIT (no dependence was detected between 295 and 80\,K). The 20\% sample shows a much weaker and actually reversed evolution for A and B peaks. Remarkably, this temperature dependence is captured by multiplet cluster calculations with the inclusion of a 1\,meV external field to represent the exchange (i.e., AF) field due to the neighboring spins. The multiplet calculations are based on parameters from our {\it ab-initio} density functional theory results \cite{hossain_PRL}, and are thus not an arbitrary fit to the data. At room temperature [red line in Fig.\,\ref{Fig1}(c,d) for 20\%], the lineshape is uniquely determined by the orbital population (at 300\,K because $T\!\gg\!H^{ex}\!=\!1$meV$\simeq\!10$K, we are effectively in a non-magnetic ground state). At lower temperatures, effects stemming from AF correlations are seen in Fig.\,\ref{Fig1}(b,d). The opposite temperature trends observed for the two Mn concentrations is reproduced by the inclusion of exchange fields with different orientations with respect to the $c$-axis (see caption). This demonstrates the extreme sensitivity of XAS to the onset of AF correlations and, more specifically, that Mn impurities do induce AF in Sr$_3$Ru$_2$O$_7$, with an exchange field orientation that varies progressively from out-of-plane to in-plane upon increasing the Mn-content from 5 to 20\%.

A summary of the temperature-dependent XAS study of 5, 10, and 20\% Mn-substituted Sr$_3$(Ru$_{1-x}$Mn$_x$)$_2$O$_7$ is shown in Fig.\,\ref{Fig1}(e), where the relative intensity change of A and B peaks [$L_2^R(T) / L_2^R(15)$, see caption for definition] is compared to the electrical resistivity data \cite{mathieu}. It is remarkable that temperature dependent XAS data can quantitatively track the trend provided by transport, which probes the charge carrier response of the system. Clearly, a sharp transition is observed in the XAS data for 5 and 10\% Mn, at $T_{MIT}\!\simeq\!50$ and 80\,K respectively, while only a continuous evolution for 20\% Mn. Together with the analysis presented in Fig.\,\ref{Fig1}(a-d),  linking the XAS temperature dependence to the onset of an AF exchange field at the Mn sites, this establishes a direct connection between the MIT  and the emergence of AF correlations. 

Whether or not the onset of this AF exchange field at $T_{MIT}$ happens simultaneously with the establishment of long-range AF order -- so that $T_{MIT}$ would be equivalent to a $T_{order}$ for some Mn concentrations -- cannot be concluded from XAS, since the latter is only sensitive to the local environment of the element being probed. 
\begin{figure}[t!]
\centerline{\epsfig{figure=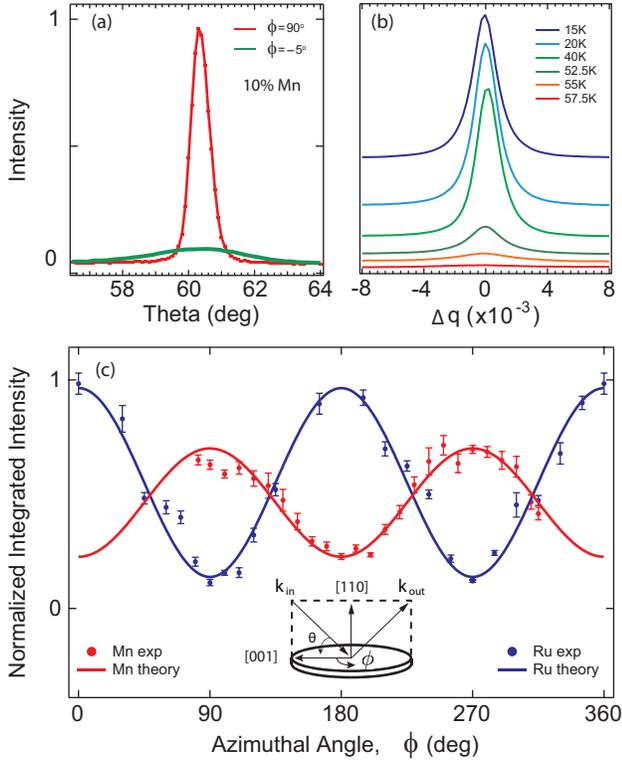,width=0.95\linewidth,clip=}} \vspace{-0.25cm}
\caption{(color online). (a) REXS transverse-momentum scan: 10\,\% Mn rocking curves ($\theta$
{\it scan}) for two different azimuthal angles measured at 641\,eV and 20\,K: $\phi=-5^\circ$ corresponds to a
$\sim\!(1/4, 1/4,l)$ scan probing the $c$-axis component of the order; $\phi=90^\circ$ corresponds to a
$(h,-k,0)$ scan probing the $ab$ in-plane component. (b) REXS longitudinal-momentum scan: 10\,\% Mn $(\frac{1}{4}+\Delta
q,\frac{1}{4}+\Delta q,0)$ scans ($\theta\!-\!2\theta$ {\it scan}) at 641\,eV for different
temperatures. (c) Azimuthal-angle dependence of the 10\,\% Mn $(\frac{1}{4}+\Delta q,\frac{1}{4}+\Delta q,0)$ momentum scan
integrated intensity, at Ru $L_2$ (2\,968\,eV) and Mn $L_3$ edges, fitted to the
formula $I^{total}_{(\frac{1}{4},\frac{1}{4},0)}\!=\!I^{\pi\rightarrow\sigma'}_{(\frac{1}{4},\frac{1}{4},0)}+
I^{\pi\rightarrow\pi'}_{(\frac{1}{4},\frac{1}{4},0)}\!\propto\!\left|\cos\theta \cos \phi
\right|^2\!+\!\left| \sin 2\theta \sin\phi \right|^2$. Due to the different Mn and Ru scattering angles
($\theta_{Mn}\!=\!61.6^{\circ}$; $\theta_{Ru}\!=\!10.9^{\circ}$), the ratio between
$(\pi\!\rightarrow\!\sigma^\prime)$ and $(\pi\!\rightarrow\!\pi^\prime)$ scattering signals is also different and the
maximum intensity angle is shifted by 90$^{\circ}$ in $\phi$.}\label{Fig2}
\end{figure}
\begin{figure}[t!]
\centerline{\epsfig{figure=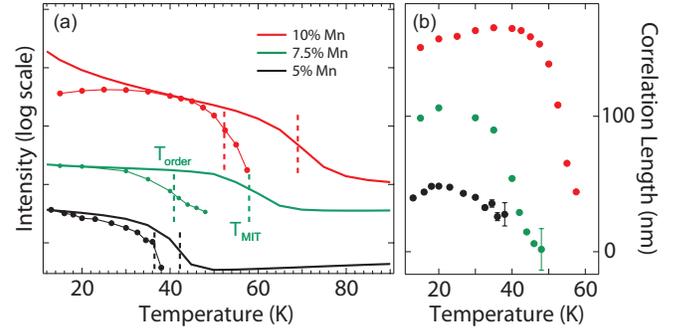,width=1.0\linewidth,clip=}} \vspace{-0.25cm}
\caption{(color online). (a) Integrated REXS momentum scans vs. temperature (connected symbols) for 5, 7.5, and 10\% Mn, together with transport data (lines), all in log scale and vertically shifted. The REXS intensity is scaled to emphasize $T_{MIT}$ and $T_{order}$ temperature difference: $T_{MIT}$ denotes the onset of the local AF exchange-field around Mn impurities, which triggers a metal-insulator transition; $T_{order}$ is the onset of magnetic order, i.e. long-range AF involving both impurity and Ru host. 
(b) REXS correlation length vs. temperature; no REXS peak was observed on 2.5\% and 20\% Mn samples.}
\label{Fig3}
\end{figure}
To address this point, and whether both Ru and Mn are participating in the order, we have done REXS studies at the Ru and Mn $L$ edges. Note that, although Mn impurities are randomly distributed, if a global order exists it is still possible to detect it because constructive interference is dependent not on the spatial location of the impurities but rather on the proper phase relation between their magnetic moments -- provided that the Mn atoms occupy substitutional Ru sites and one is indeed probing magnetic scattering. Hence, the existence of superlattice peaks at the Mn $L$-edge, in addition to those detected at the Ru $L$-edge, is a direct confirmation of spin correlation between Mn impurities. Fig.\,\ref{Fig2}(a) shows the 10\,\% Mn $L_3$-edge $(\frac{1}{4},\frac{1}{4},0)$ rocking curves at $T\!=\!20$\,K, i.e. the $\theta$-angle dependence of the intensity at 641\,eV for different azimuthal angles $\phi$ (here we focus on the $L_3$-edge since it is the most intense \cite{hossain_PRL} and thus better suited for REXS). Since the peak width in momentum is proportional to the inverse of the correlation length, the sharp $q_{x,y}$ dependence ($\theta$ scan at $\phi\!=\!90^\circ$) and the broad response in $q_z$ ($\theta$ scan at $\phi\!=\!-5^\circ$) indicate a two-dimensional order with weak correlation along the $c$-axis. The structurally forbidden reflections are detected only below $T_{MIT}$, with a progressively increasing strength upon reducing temperature (and increasing Mn concentration, as discussed below). This is shown for 10\,\% Mn substitution in Fig.\,\ref{Fig2}(b), by the Mn-$L_3$ edge $(\frac{1}{4}+\Delta q,\frac{1}{4}+\Delta q,0)$ scan at 641\,eV. 

Most important, it should be noted that: ({\it i\,}) the order is predominantly electronic and not structural, since we did not observe corresponding superlattice reflections in non-resonant low-temperature x-ray diffraction; ({\it ii\,}) it is also not associated with the spatial ordering of the Mn impurities, as evidenced by the Mn-concentration independence of the REXS $q$-vectors in the range $0.05\!\leq\!x\!\leq0.10$. Altogether, this points at an incipient electronic instability in the Ru host, which is simply triggered by the Mn impurities. 
\begin{figure}[t!]
\centerline{\epsfig{figure=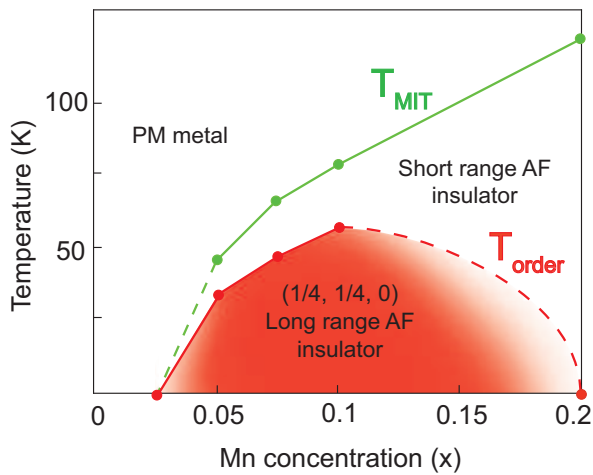,width=0.9\linewidth,clip=}} \vspace{-0.25cm}
\caption{(color online). Sr$_3$(Ru$_{1-x}$Mn$_x$)$_2$O$_7$ phase diagram: the green line is the MIT phase boundary as seen by transport   \cite{mathieu} and XAS (Fig.\,\ref{Fig1}), which we interpret as onset of AF short-range correlations; the red line defines the onset of long-range spin-order between the Mn impurities via the quasi ordered Ru-O host (dashed sections are the boundaries suggested by experimental points on either side, although not directly verified). Our results indicate the region between green and red lines to be a Mn-driven disordered AF insulating state.}\label{Fig4}
\end{figure}
The nature of this electronic order (i.e. spin, charge, or orbital) can be further clarified by the detailed azimuthal-angle dependence of the integrated intensity of the Ru and Mn-edge $(\frac{1}{4},\frac{1}{4},0)$ peaks. The results for the 10\,\% Mn system are presented in Fig.\,\ref{Fig2}(c), together with a theoretical angle-dependence calculated for pure spin order \cite{Hill}. Although a charge component cannot be completely excluded, the agreement between measured and calculated azimuthal dependence -- consistent with XAS as well as magnetic susceptibility results \cite{Hu} -- implies the primarily magnetic nature of the order.

The results in Fig.\,\ref{Fig2} reveal the presence of a two-dimensional {\bf Q=}$(\frac{1}{4},\frac{1}{4},0)$ long-range spin order below T$_{MIT}$; next we will address at which temperature exactly does the order emerge and its dependence on Mn-content. Fig.\,\ref{Fig3}(a) shows the Mn $L_3$-edge REXS integrated scattering peak intensity as a function of temperature, together with the transport data, for 5, 7.5, and 10\% Mn substitution. It is striking that the scattering peak intensity rises on a temperature scale $T_{order}$ lower than $T_{MIT}$ at all Mn contents, as emphasized in Fig.\,\ref{Fig3}(a) by the vertical dashed lines ($T_{order}$ is defined as the temperature at which the peak intensity reaches half of its maximum value). For a deeper understanding of the difference between the onset of Mn-site exchange field at $T_{MIT}$ and of long-range antiferromagnetism at $T_{order}$, it is helpful to have a closer look at the temperature dependence of the REXS correlation length in Fig.\,\ref{Fig3}(b), which provides an estimate for the average size of the magnetic islands. The correlation length gradually increases upon lowering the temperature, before saturating and eventually decreasing, as typically seen in reentrant spin glasses due to the interplay of competing interactions and disorder \cite{Maletta,Aeppli}. For the 7.5\% Mn concentration we can reliably follow the REXS peaks down to a very short correlation length of $\sim$6\,nm at 46\,K. This is equivalent to an area of $\sim$15$\times$15 unit cells, which would contain $\sim$17 Mn impurities. Hence, between 46\,K and $T_{MIT}^{7.5\%}\!\simeq\!57$\,K, at which the Mn-site AF exchange field first appears, the correlation length spans less than 17 impurities and magnetic correlations are indeed short range. However, between 46 and 36\,K, the correlation length rises from 6 to 90\,nm, i.e. from $\sim$17 to 3800 Mn sites. This demonstrates that $T_{MIT}$ in Mn-Sr$_3$Ru$_2$O$_7$ is not the onset of AF order but rather of AF islands. The MIT is thus of the Mott-type, in which electronic correlations drive both mass enhancement and corresponding transport anomaly, as well as the onset of local AF correlations. In the range $T_{order}\!<\!T\!<\!T_{MIT}$, magnetic correlations between separate Mn impurities are weak; true long-range AF order appears only below $T_{order}$, once the correlation length extends over a critical number of Mn impurities. 

Our results can be summarized into the phase diagram of Fig.\,\ref{Fig4}; this is consistent with the one in Ref.\,\onlinecite{Hu}, with in addition the conclusive identification of the susceptibility maximum $T_M$ line with $T_{order}$, i.e. the boundary  of a bona-fide long-range AF phase. As for the detailed evolution of the $(\frac{1}{4},\frac{1}{4},0)$ magnetic order, it should be emphasized that no REXS superlattice peaks were detected for 2.5\% and 20\% Mn-substituted samples: the long-range AF order appears between 2.5 and 5\% Mn, disappearing again upon approaching 20\%. The lower limit of the ordered phase stems from the insufficient overlap between the AF patches surrounding the Mn impurities for $x\!<\!0.05$. As for the upper limit, one should note that the 100\% substituted compound, i.e. Sr$_3$Mn$_2$O$_7$, is  G-type AF insulator with $T_N\!\simeq\!160$\,K \cite{mitchell}; this would lead to a competition between $(\frac{1}{4},\frac{1}{4},0)$ and G-type orders -- and eventually the establishment of the latter -- as the Mn content increases. Finally, since Sr$_3$Ru$_2$O$_7$ does not show any long-range magnetic order, it is clear that the latter is induced by the $S\!=\!2$, 3$d$-Mn$^{3+}$ impurities \cite{mathieu,hossain_PRL,Panaccione_NJP}. Nevertheless, the ordering is independent of the precise 5-10\,\% Mn concentration, suggesting that the role of Mn is primarily that of triggering and/or stabilizing a tendency already incipient in the parent compound.

We thank M.Z. Hasan for the use of the ALS scattering chamber. This work was supported by the Max Plank - UBC Centre for Quantum
Materials, the Killam, Sloan, and NSERC's Steacie Fellowship Programs (A.D.), CRC (A.D. and G.A.S.), ALS (M.A.H), NSERC, CFI, CIFAR Quantum Materials, and BCSI. ALS is supported by the U.S. DOE Contract No. DE-AC02-05CH11231.

\vspace{-0.4cm}

\bibliographystyle{plain}

\end{document}